\documentclass[12pt]{article}
\usepackage[top=1in, bottom=1in, left=1in, right=1in]{geometry}
\usepackage{setspace}
\usepackage{graphicx} 
\usepackage{amsmath} 
\usepackage{caption} 
\usepackage{fancyhdr} 
\usepackage{titling} 
\usepackage{datetime} 
\usepackage{float} 

\usepackage{ulem} 
\usepackage{xcolor} 

\usepackage[T1]{fontenc}
\DeclareSymbolFont{TOneChars}{T1}{\familydefault}{m}{it}
\SetSymbolFont{TOneChars}{bold}{T1}{\familydefault}{bx}{it} 
\DeclareMathSymbol{\mathdh}{\mathord}{TOneChars}{"F0}

\renewcommand{\maketitle}{
    \begin{titlepage}
        \centering
        \vspace*{1cm}
        {\bfseries \MakeUppercase{Gravitational Wave Propagation through Viscous Matter}\par}
        \vspace{1.5cm}
       { Vishnu Kakkat\textsuperscript{1,5}\textsuperscript{a}, Ulrich K. Beckering Vinckers\textsuperscript{2}\textsuperscript{b}, Nigel T. Bishop\textsuperscript{2,5}\textsuperscript{c}, \\ Amos S. Kubeka\textsuperscript{3,5}\textsuperscript{d}, Monos Naidoo\textsuperscript{2,5}\textsuperscript{e}, Udaykrishna Thattarampilly\textsuperscript{4}\textsuperscript{f}, \\ and Petrus J. van der Walt\textsuperscript{2}\textsuperscript{g} \\
        \vspace{.5cm}
        \textsuperscript{1}Institutionen för teknikvetenskap och matematik, Luleå tekniska universitet, 971 87 Luleå, Sweden\\
        \textsuperscript{2}Department of Mathematics, Rhodes University, Makhanda 6140, South Africa\\
        \textsuperscript{3}Department of Mathematical Sciences, University of South Africa, Pretoria, South Africa\\
         \textsuperscript{4}Center for Gravitation and Cosmology, College of Physical Science and Technology, Yangzhou University, Yangzhou 225009, China\\
         \textsuperscript{5}National Institute for Theoretical and Computational Sciences (NITheCS), Stellenbosch, South Africa}\\
        \vspace{1cm}
        Emails: \textsuperscript{a} vishnu.kakkat@associated.ltu.se, \textsuperscript{b} ulrich.beckeringvinckers@ru.ac.za, \\\textsuperscript{c} n.bishop@ru.ac.za, \textsuperscript{d} kubekas@unisa.ac.za, \textsuperscript{e} monos.naidoo@ru.ac.za, \\\textsuperscript{f} uday7adat@gmail.com, \textsuperscript{g} peetvdw@worldonline.co.za\\
        \vspace{0.5cm}
        Corresponding author: vishnu.kakkat@associated.ltu.se
                \vspace{1cm}
        \begin{abstract}
         
It has been known that gravitational waves (GWs) transfer energy to viscous matter through which they propagate, but the effect is too weak to be astrophysically significant. Using linearized perturbations about a Minkowski background, we previously showed that the interaction can become important when the distance between matter and source is smaller than the GW wavelength. Here, we review extensions to more realistic backgrounds, namely Schwarzschild spacetime and a static spherically symmetric setting. We find that GW damping and the associated heating of the viscous fluid are enhanced, and can lead to substantial attenuation or even gamma-ray bursts. We investigate astrophysical scenarios where these effects may be relevant, including core-collapse supernovae, binary neutron star mergers, and accretion onto binary black hole mergers. {Our analysis models only the GW damping and heating effects within linear perturbation theory on prescribed background spacetimes; thus, the quantitative results should not be interpreted as precise predictions for fully dynamical astrophysical systems, but rather as identifying scenarios in which these effects may be important.}
        \end{abstract}
           \end{titlepage}
}

\begin{document}

\maketitle

\section{Introduction}
\label{intro}
It has been known since 1966~\cite{hawking1966perturbations} that when gravitational waves (GWs) propagate through viscous matter, there is a transfer of energy from the GWs to the matter. In the case of plane, linearized GWs propagating a distance $\Delta$ in the $z$-direction through a fluid with shear viscosity $\eta$, the GW strain obeys
\begin{equation}
H(z+\Delta)=H(z)\exp\left(-\frac{8\pi G\eta \Delta}{c^3}\right),
\label{e-dHdz}
\end{equation}
where $H$ is the magnitude of the GW. In SI units, $G/c^3={\mathcal O}(10^{-36})$, and there are no known astrophysical or cosmological circumstances in which $\eta\Delta$ approaches values large enough for this effect to be significant. Consequently, it has become common practice to regard GWs as passing transparently through matter.

Using the Bondi–Sachs formalism~\cite{Bondi:1962px,Sachs:1962wk}, we have gone beyond the plane wave approximation and obtained expressions for GW damping and the consequent heating of the viscous fluid. Initially, this was done for perturbations about Minkowski spacetime surrounded by a spherical shell of matter~\cite{bishop2022effect,Naidoo:2021rzw}. In the regime where the GW wavelength $\lambda$ is much smaller than the distance $r$ from the source to the matter, Eq.~\eqref{e-dHdz} is recovered. However, when $\lambda/r \gg 1$, the damping effect can become substantial, and there exist scenarios in which it may be astrophysically significant.

Linear perturbations about Minkowski spacetime have the advantage that the metric can be expressed in terms of elementary functions, but the applicability of such a model to realistic astrophysical systems is limited. We therefore consider a Schwarzschild background~\cite{Bishop:2024htv}, and subsequently a general, non-vacuum, static, spherically symmetric background spacetime~\cite{Bishop2026spherical}. The previous separation-of-variables ansatz is retained, reducing the Einstein equations to a system of ordinary differential equations in the radial coordinate $r$. However, in the more general setting, the resulting system must be solved numerically. This is technically challenging because the ODEs possess an essential singularity as $r\to\infty$. It is found that, relative to the Minkowski case, the modified background enhances the GW damping and heating by a factor that can reach ${\mathcal O}(10)$ in the applications considered.

The astrophysical scenarios investigated include core-collapse supernovae (CCSNe), binary neutron star mergers (BNS), and the presence of matter during a binary black hole merger (BBH). There remain uncertainties in the relevant astrophysical parameters, as discussed later in this work; the following statements refer to representative mid-range values. For both CCSNe and BNS systems, the GW damping is predicted to be nearly complete. The associated temperature increases are significant and are expected to influence the overall evolution of these systems, although detailed modeling lies beyond the scope of this paper. In contrast, for a BBH system accreting matter from an external source (an unlikely configuration), the damping of the GWs would be minimal. However, the temperature increase could be sufficient to generate a gamma-ray burst, possibly related to the observation reported in~\cite{connaughton2016fermi}.

The essay is organized as follows. Section~\ref{sec:theory} introduces the Bondi--Sachs formalism and the thin-shell model in Minkowski spacetime. Section~\ref{sec:thin} discusses the Schwarzschild background, and Section~\ref{sec:gen_thin} extends the model to a general static spherically symmetric background. Section~\ref{sec:astrop} presents the astrophysical applications, and the final section concludes.
\section{Theoretical formulations}
\label{sec:theory}

The Bondi--Sachs (BS) formalism \cite{Bondi:1962px,Sachs:1962wk}, introduced in 1962, is the most natural framework for studying GW radiation. In this formalism, spacetime is foliated into outgoing null hypersurfaces, labeled by retarded time $u = t - r/c$. Each such surface is generated by the null rays emitted from the source at a given moment $u$, and the coordinates follow those
rays outward. These hypersurfaces are the wave fronts of the GW radiation.
 
\subsection{The Bondi--Sachs metric}

The Bondi--Sachs metric in null coordinates takes the following form 
\begin{eqnarray}
ds^2=&-&\left[e^{2\beta}\left(1+\frac{W}{r}\right)-r^2h_{AB}U^{A}U^{B}\right]du^2-2e^{2\beta}dudr\nonumber\\
     &-&2r^2h_{AB}U^{B}dudx^{A}+r^{2}h_{AB}dx^{A}dx^{B},
\label{eq:BS_metric}
\end{eqnarray}
where $x^A $ are the angular coordinates, $r$ is the surface-area radial
coordinate satisfying $\det(h_{AB}) = \det(q_{AB})$ with $q_{AB}$ the unit sphere metric,
and $h_{AB}$ is the conformal angular metric with only two independent degrees of freedom.

All four metric functions are of physical significance. The function $\beta$
controls the gravitational redshift along a null ray. The function $W$ carries the energy content of
the spacetime; at large $r$ its $1/r$ coefficient gives rise to the Bondi mass, the total
energy of the system. The complex field $U$, formed by contracting $U^A$ with the complex
angular dyad $q^A$, measures how null rays are tilted relative to the angular grid and
vanishes for spherical symmetry. The complex spin-weighted-2 field $ J = \tfrac{1}{2}\,h_{AB}\,q^A q^B$,
 measures how the two-sphere of null rays deviates from a perfect sphere. The two
independent components of $J$ correspond to the two polarisation modes
of the gravitational radiation. A non-zero $J$ is the geometric signature of a GW: it
stretches the null sphere in one direction and compresses it in the orthogonal one. All
functions $\beta$, $W$, $U$, $J$ vanish identically in a flat Minkowski spacetime.

\subsection{ Einstein equations}
In the BS formalism, the Einstein equations can be organized into a clean hierarchy\cite{Naidoo:2021rzw}, namely,
\begin{itemize}
\item the hypersurface equations $R_{rr}$, $q^{A}R_{rA}$, $h^{AB}R_{AB}$ for $\beta$, $U$, and $W$ respectively,
\item the evolution equations $q^{A}q^{B}R_{AB}$ for $J$, and
\item the constraints equations $R_{0a}$.
\end{itemize}
A procedure to solve the above equation has been studied previously. A summary of the procedure is as follows. The hypersurface equations integrate sequentially outward along each null ray to determine $\beta, U$, and $W$, all entirely by radial integration. 
Once these are known, the evolution equation advances $J$ forward, propagating the radiation in retarded time. The remaining equations are constraints that are automatically satisfied once the other equations are satisfied, serving as consistency checks. 
\subsection{Spherical thin shell and GW damping}

\begin{figure}[h]
  \centering
  \includegraphics[width=0.45\textwidth]{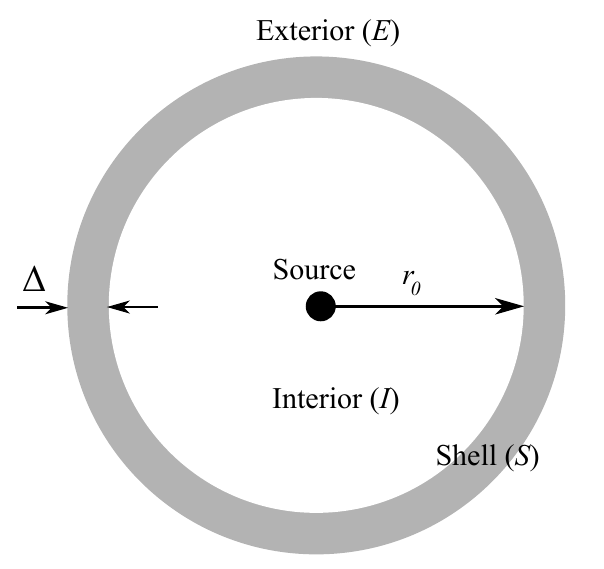}
  \caption{A GW source at the origin surrounded by a spherical matter shell
           of thickness $\Delta$ located between radii $r_0$ and $r_0+\Delta$.
           The spacetime is divided into three regions: interior (I), shell (S),
           and exterior (E). The BS equations are solved analytically in each
           region and matched across the boundaries.}
  \label{fig:shell}
\end{figure}

We will now consider a scenario where a GW source is surrounded by a thin spherical shell of viscous matter in a Minkowski background (see Figure \ref{fig:shell}). 
The linearized BS equations are solved separately in each of the three regions, and the solutions are matched across the shell boundaries by requiring that the metric functions are continuous 
\cite{bishop2022effect,Bishop2024}. The interesting feature of this solution is its near-field corrections. Beyond the familiar $1/r$ term that dominates at large distances 
the solution contains faster-decaying contributions at orders $1/r^2$, $1/r^3$, and higher. When a source is surrounded by a viscous fluid, these near-field corrections to the structure of the GWs will also contribute to the damping of GWs, giving a dissipation rate that exceeds the standard far-field estimates. 

It has been shown that the GW energy dissipated on the viscous shell of matter is given by
\begin{equation}\label{eq:Energy_m}
		\left<\dot{E}_\eta\right> = -16\pi\eta \frac{G}{c^3} \delta r \left<\dot{E}_{GW}\right>
		\left(1+\frac{2}{r^2\nu^2}+\frac{9}{r^4\nu^4}
		+\frac{45}{r^6\nu^6}+\frac{315}{r^8\nu^8}
		\right),
  \end{equation}
where {$\nu$ is the angular frequency of the GW,} $\delta r$ represents the thickness of the shell and $\left<f\right>$ represents the average of $f$ over the wave period.
The rate of energy loss $\left<\dot{E}_\eta\right>$ is calculated using the 
expression $-2\eta \sigma^{ab}\sigma_{ab}$, where $\sigma_{ab}$ is the shear tensor.  The negative sign shows the flow of energy from the GWs to the viscous shell.
Due to the dissipation of energy caused by  GWs, two phenomena may occur. These are the damping of GWs and the associated heating of the viscous shell.

The GW amplitude decays as
\begin{equation}
H(r_o)=H(r_i)\exp\left(-\frac{45\eta\lambda^8}{32r_i^7\pi^7}\right),
\label{h2}
\end{equation}
showing that damping can be strong when the source--matter separation is small compared to the wavelength. The associated temperature increase is
\begin{equation}
\Delta T= \frac{\sqrt{\pi}G\eta}{6c^5C\rho}\nu^2\Delta E_{GW}D,
\label{eqn:diffT}
\end{equation}
where $\rho$ is the density, $C$ the specific heat capacity, and $D$ is a function of frequency of GW and radius of the shell. 

{At this point, we note that it remains unclear how significantly the GW damping effects reviewed here would be modified when taking into account higher-order perturbations. Despite this limitation, the linearized results studied here provide us with insight into the damping mechanisms that arise in the context of GR. Furthermore, the application of higher-order perturbation theory presents additional difficulty, particularly in treating terms involving the mixing of radiative and non-radiative contributions. While such mixing at the level of the perturbed metric is beyond the scope of the present work, it is possible to consider the effect of including a non-radiative contribution in the background metric. This is accomplished by taking the background metric to be Schwarzschild instead of Minkowski, which is the subject of the following section.}
\section{Thin  viscous shell in Schwarzschild background}
\label{sec:thin}

We next extend the Minkowski calculation to a Schwarzschild background~\cite{Bishop:2024htv}. While the field equations linearized around a Minkowski background can be solved analytically as discussed above, extending such a study to a background that corresponds to the Schwarzschild solution requires the use of numerical methods. In particular, the master equation obtained through the manipulation of the $q^A R_{Ar}$ and $q^A q^B R_{AB}$ equations (see, for example,~\cite{Bishop:2004ug}) is then recast as a Riccati-type equation and solved numerically. This is done after first obtaining suitable initial data by substituting a series expansion into the original master equation. Further details regarding the numerical implementation can be found in~\cite{Bishop:2024htv}.

\begin{figure}
    \centering
    \includegraphics[width=0.6\linewidth]{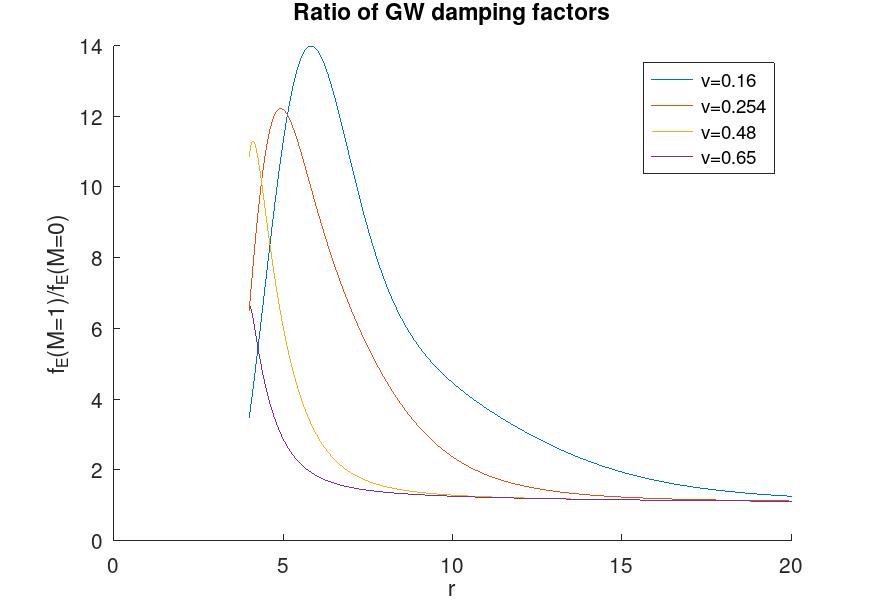}
    \caption{Profile plots based on Ref.~\cite{Bishop:2024htv} showing the ratio of $f_E(r, \nu, M)$ for the Schwarzschild ($M=1$) and Minkowski ($M=0$) cases for fixed values of $\nu$.} 
    \label{fig:damping_ratio}
\end{figure}

Once the metric components have been obtained, one can consider a matter shell with four-velocity $V_a$ which is normalized to unity. The normalization condition can be used to find an expression for $V_0$, while the conservation equation can be used to obtain $V_1$ and $q^A V_A$. Following the same procedure as in~\cite{bishop2022effect}, we obtain a similar expression for the energy dissipation by GW as
\begin{align}
    \langle \dot{E}_s \rangle = -16 \pi \eta \frac{G}{c^3} \delta r \langle \dot{E}_{GW} \rangle f_E(r, \nu, M) \,.
\end{align}
A comparison was done by plotting the ratio of $f_E$ in Figure~\ref{fig:damping_ratio} for the cases of a Minkowski background and a Schwarzschild background, i.e., for $M = 0$ and $M = 1$. 

By noting that $\nu = 2\pi M/\lambda$, for $M\ne 0$ and where $\lambda$ is the GW wavelength, it is evident from Figure~\ref{fig:damping_ratio} that the difference between the two cases is minimal when $r\gg M$ or $\lambda < M$. It is also worth considering examples where the difference between the two cases is significant. One such example is the case where $\nu = 0.254$ which, as noted in~\cite{Bishop:2024htv}, corresponds to the merger frequency of GW150914~\cite{abbott2016observation}. For such a choice of $\nu$, Figure~\ref{fig:damping_ratio} demonstrates that the damping and heating effects are about 9 times greater in the Schwarzschild case when $r \sim 6M$ and $\lambda \sim 25 M$.

{In the set-up considered in this section, the primary non-radiative contribution from the source is included in the background metric. An alternative approach would be to keep the background as being Minkowski, and include the non-radiative part in the perturbed metric. While these two approaches should agree when taking into account the smallness conditions associated with the linearized regime, the results obtained using these two approaches when such conditions are violated may not coincide. Even with this ambiguity, the analysis reviewed above indicates that there may be astrophysical scenarios where the difference between considering a Schwarzschild background and a Minkowski one is significant.}

\section{Thin viscous shell in a general spherically symmetric setting}
\label{sec:gen_thin}

We further extend the previous results to a general static, spherically symmetric background
containing a thick shell of viscous matter exterior to a GW source. The unperturbed configuration consists of an inner core of mass $M_c$ and radius $r_c$. A static fluid shell occupies
$r_c<r<r_t$ and is characterized by density and pressure. For $r>r_t$, the spacetime is
vacuum Schwarzschild with total mass $M_t$ enclosed within $r<r_t$.

\begin{figure}
\includegraphics[scale=0.28]{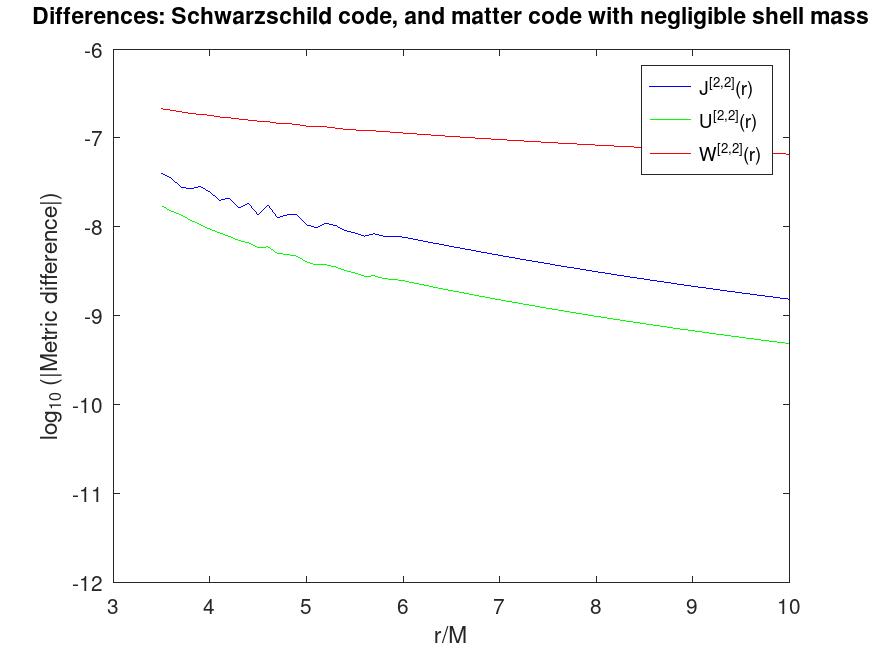}
\includegraphics[scale=0.28]{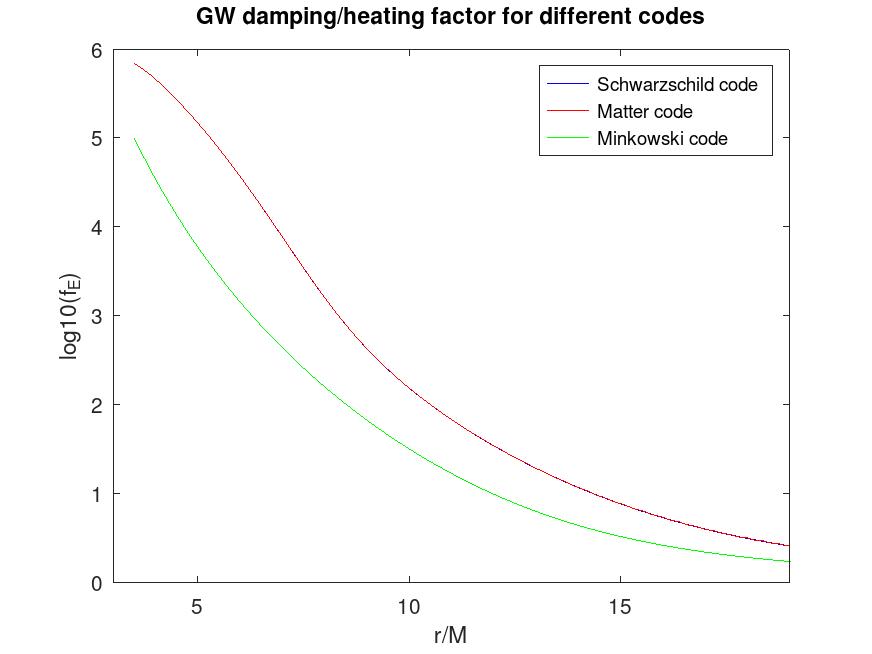}
\caption{{Comparison of the Schwarzschild, Matter, and Minkowski codes. The parameter $\nu = 0.13998$, corresponding to a frequency of 74.9Hz when $M=60M_\odot$, is used. The left panel shows, on a $\log_{10}$ scale, the differences between the indicated metric quantities computed using the Schwarzschild and Matter codes. The right panel shows $\log_{10}(f_E)$ obtained using the Schwarzschild and Matter codes, together with the corresponding Minkowski-background result from Eqn.~\eqref{eq:Energy_m}.}}
\label{f-Spert}
\end{figure}

{The same technique as in the Schwarzschild case is employed, with details of the numerical implementation given in Ref.~\cite{Bishop:2024htv}. In what follows, we refer to the calculations based on perturbations about Minkowski and Schwarzschild backgrounds as the Minkowski code and Schwarzschild code, respectively, while the numerical implementation for a general static, spherically symmetric matter background is referred to as the Matter code. We obtain numerical solutions with errors below $10^{-14}$
 in all cases. As a consistency check, Figure~\ref{f-Spert} compares the Matter code in the limit of negligible shell density with the Schwarzschild code, while also showing the corresponding Minkowski-background result for the GW damping/heating factor $f_{E}$.}
Since the metric is constructed numerically, the damping and heating effects of GWs are also calculated numerically and applied to astrophysical settings.

\section{Astrophysical applications}
\label{sec:astrop}

\subsection{Overview}

In this section, we consider three astrophysical scenarios that can be framed, with some assumptions, as thin shell models. These are: Core-collapse supernovae (CCSNe), binary neutron star mergers (BNS) and the accretion of matter surrounding binary black hole mergers. The scenarios were specifically selected since the resulting GWs of these events are within the detection capabilities of current GW observatories. Detailed analysis of these were presented in \cite{Bishop2024} using perturbations on a Minkowski background and, it was repeated using perturbations on a Schwarzschild background in~\cite{Bishop:2024htv} and further extended in~\cite{Bishop2026spherical}. 

\subsection{Core-collapse supernovae}

Core-collapse supernovae (CCSNe) represent one of the most violent astrophysical processes in the Universe. As the iron core of a massive star exceeds the Chandrasekhar mass {($\sim 1.4\,M_{\odot}$)}, electron degeneracy pressure fails, triggering gravitational collapse \cite{Bethe:1990mw,Janka:2012wk,Jerkstrand:2025bea}. The collapse halts when nuclear densities ($\rho \sim 2-3 \times 10^{17}\,\mathrm{kg\,m^{-3}}$) are reached, leading to the formation of a proto-neutron star (PNS), sudden rebound, bounce and an ensuing outward shock wave. The post-bounce hydrodynamics and anisotropic mass motions produce gravitational waves expected to be detectable by current observatories \cite{Yakunin:2017tus,Andresen:2016pdt}.
	
More precisely, in terms of time scales, the CCSNe evolution proceeds as follows: The collapse occurs on a timescale of $\sim 200$--$400\,\mathrm{ms}$ from onset to core bounce \cite{Muller:2020ard}. Then, the \textit{bounce} occurs within $\sim 1\,\mathrm{ms}$, producing a sharp burst of gravitational radiation \cite{Ott:2012jq}. Following the bounce, convection and the standing accretion shock instability (SASI) drive stochastic mass motions and asymmetric neutrino emission, both contributing to longer-duration GW signals lasting up to several hundred milliseconds \cite{Kuroda:2016bjd,Andresen:2018aom}.

The above translates to the following frequency bands, which depend strongly on the mass and rotation of the progenitor: Bounce signals ranging between $\sim 500$--$1000\,\mathrm{Hz}$ with a duration of $\sim 10\,\mathrm{ms}$. Convection/SASI ranging between $\sim 100$--$300\,\mathrm{Hz}$, lasting $\sim 300\,\mathrm{ms}$. With a PNS $g$-mode oscillations range of $\sim 1$--$2\,\mathrm{kHz}$.

The GW strain amplitude from a CCSN at 10\,kpc is expected to be $H \sim 10^{-22}$--$10^{-21}$, peaking near bounce and modulated by subsequent PNS oscillations \cite{Yakunin:2017tus}.

Using Eqs. (17) and (18) in  \cite{Bishop2026spherical} along physically realistic parameter values (see Table I in \cite{Bishop2026spherical}) and varying the frequency over $100$--$1000~\textrm{Hz}$ and the viscosity over $10^{23}$--$10^{25}~\mathrm{kg\,m^{-1}\,s^{-1}}$. For these computations, the total mass is $1M_\odot$, the core mass is $0.71M_\odot$ and the core radius is $10$km. The GW damping factor is reported at the outer boundary and the temperature rise at the inner boundary. Results from the computations are given in \cite{Bishop2026spherical}, where previous values for the Minkowski background are included for comparison.

\begin{figure}[H]
	\begin{center}
    \includegraphics[scale=0.6]{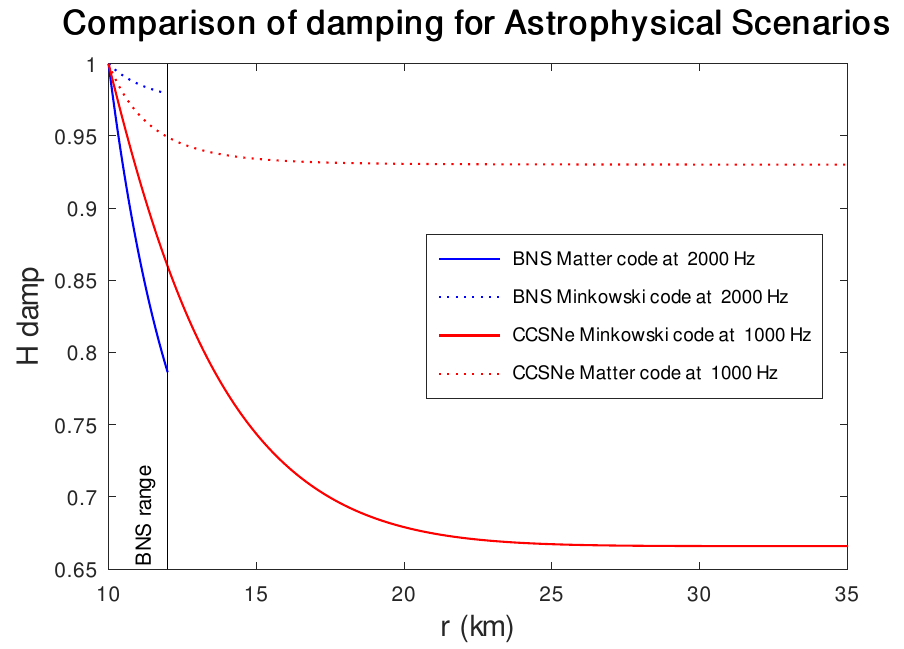}
    \caption{H damp plotted against distance (km) for the CCSN and BNS models.}
    \label{f-CCSN_BNS}
	\end{center}
\end{figure}

The results obtained have several implications, which include: First, it is important to use the matter code, since results obtained using the Minkowski code always underestimate the GW damping and heating effects (see Figure~\ref{f-CCSN_BNS}). Second, in all models, the GW damping was almost complete. This implies that the GW signal from a CCSNe, even if nearby, would not be observable except possibly in a frequency band greater than $1$kHz. Third, the predicted temperature increases need to be compared to the ambient temperature in the mantle of a CCSNe, which is about $10^{12}$K \cite{boccioli2024physics}, and this value is exceeded in the lower frequency band ($100$Hz). However, the temperature increase occurs deep within the star and is not directly observable. It may be that it could lead to observable effects, but the required modeling is beyond the scope of this paper.
\subsection{Binary neutron star mergers}

From the first detection of GWs from the binary neutron star merger GW170817 to subsequent events, BNS mergers have been recognized as major GW sources. The parameter estimates presented  in \cite[Table III]{Bishop2026spherical} correspond to the post-merger stage of a BNS system, based on numerical and observational results reported in~\cite{abbott2017gw170817,abbott2017multi,abbott2020gw190425,abbott2017gravitational,hammond2025investigating,fonseca2021refined,antoniadis2013massive,fontbute2025gravitational}. Moreover, these estimates are consistent with those inferred from GW170817~\cite{abbott2019properties,abbott2016observation,abbott2017gw170817,abbott2017gravitational,abbott2020gw190425}. 
Considering both long-lived and short-lived remnants, the core mass lies in the range \(2.4\text{--}2.8~M_\odot\), with a corresponding radius of \(10\text{--}12~\mathrm{km}\). Including the surrounding accretion disk, the total bound mass is \(2.7\text{--}3.1~M_\odot\), with the disk contributing between \(5\times10^{-4}\) and \(0.3~M_\odot\). The density of the shell is modeled as
\begin{equation}
    \rho(r) = \rho_0 \, e^{-r/r_0},
\end{equation}
where \(\rho_0 = 10^{17}~\mathrm{kg\,m^{-3}}\) and \(r_0 = 12~\mathrm{km}\).

We evaluate Eqs.(17) and (18) in  \cite{Bishop2026spherical} using physically realistic parameter values (see Table III in \cite{Bishop2026spherical}) and varying the frequency over $1-2~\textrm{kHz}$ and the viscosity over $10^{24}$--$10^{30}~\mathrm{kg\,m^{-1}\,s^{-1}}$.
{We note that numerical-relativity simulations employing 
high-resolution shock-capturing schemes possess an effective numerical viscosity 
that may exceed estimates of the physical viscosity and decreases with increasing 
grid resolution~\cite{rembiasz2017measurements}. However, numerical viscosity cannot be directly identified with 
the physical shear-viscosity coefficient $\eta$ entering our model. Nevertheless, 
existing resolution studies of binary neutron-star mergers have not reported a 
systematic resolution-dependent suppression of the GW amplitude of the type that might be expected from strong viscous damping \cite{zappa2023binary}. This provides an important consistency consideration for our predictions. Therefore, the large damping obtained at the upper end of the viscosity range considered here should be interpreted cautiously until it can be tested against fully dynamical numerical-relativity simulations with explicitly modelled physical shear viscosity.}

For these computations, the total mass is $2.7M_\odot$, the core mass is $2.64M_\odot$ and the core radius is $10$km. The GW damping factor is reported at the outer boundary and the maximum temperature rise at the inner boundary.

There are some interesting implications following from these calculations (see \cite{Bishop2026spherical} for more detailed results): First, the general trend is clear: general spherically symmetric background with high values of viscosity and the frequency exhibit stronger damping and correspondingly larger heating than the Minkowski background (see Figure \ref{f-CCSN_BNS}). Second, higher viscosity and lower GW frequency enhance the damping, thereby reducing the fraction of the wave that escapes the remnant system. This implies a reduced detectability for high-viscosity environments, due to strong GW–matter interactions. Furthermore, it has been reported in the literature that the ambient thermal state of a post-merger BNS system is expected to rise to $10-100~\textrm{MeV}$ ($\approx 10^{11}$--$10^{12}$K), primarily due to shock heating~\cite{raithel2021realistic,baiotti2017binary,paschalidis2017rotating}. Remarkably, several cases reported in~\cite[Table IV]{Bishop2026spherical} show that GW-induced viscous heating alone can reach comparable temperatures, implying that dissipation of the post-merger GW signal may contribute non-negligibly to the thermal budget of the remnant.

\subsection{Accretion at binary black hole mergers}

Binary black hole mergers currently make up the majority of detected GW sources and while these are expected to occur in an environment devoid of significant amounts of matter, occasions where a merger takes place in an environment where sources of matter can be present are also possible. In fact, on the upper scale, supermassive black hole mergers are expected to occur in the presence of significant amounts of accreting matter. In what follows, we consider a scenario where a merger with the same properties as GW150914 occurs while surrounded by matter.

The effect of an accretion disk at a binary black hole merger was considered in~\cite{bishop2024heating} and assumed a Minkowski background. As mentioned, we model the effect based on the parameters of GW150914 \cite{scientific2016tests}
\begin{equation}
\Delta E_{GW}=3M_\odot\,,
f=155 \mbox{Hz}\,,
M_f=62 M_\odot\,,
\end{equation}
where $f$ is the frequency at merger and $M_f$ is the final mass. The energy loss $\Delta E_{GW}$ is for the whole inspiral, and we use instead $2M_\odot$ ($=3.6\times 10^{47}$J) which was the energy radiated away as the frequency increased from 90Hz through peak emission at 132Hz and towards 220Hz as merger gave way to ringdown. For the purposes here, we fixed frequency to the middle of the range, i.e., 155Hz, which was found to be representative of the range and simplified the calculations.
Parameters of a stationary accretion model, as outlined in ~\cite{shakura1976theory,arai1995accretion,abramowicz2013foundations} are used to model the matter in the accretion disk: at the ISCO (Innermost Stable Circular Orbit), the dynamical viscosity $\eta$ is $3.5\times 10^9$J sec/m${}^3$, the density is $\rho$ as $4$kg/m${}^3$ and the specific heat is $1.43\times 10^4$J/kg/${}^\circ$K. The radius of the ISCO is taken as $r=329$km, being the value for a prograde orbit around a black hole of mass $M_f$ with angular momentum parameter $a=0.67$ (which is the estimated value for the GW150914 remnant).

Since the mass of the accretion disk is insignificant compared to $62M_\odot$, the GW damping and heating effects are computed using the code for a Schwarzschild background; for comparison purposes, results for the Minkowski background are also shown. It was found that GW damping is negligible, i.e., $({H_i-H_o})/{H_i} <10^{-14}$,
but that the temperature increases can be highly significant. Evaluated at an ISCO radius of $r=329$km, the Minkowski model reached $3.47\times 10^8$K at the equator compared to the Schwarzschild model reaching $2.46\times 10^{12}$K. At the poles, the Minkowski model reached $1.75\times 10^6$K and the Schwarzschild model~{$3.77\times 10^6$K.}

Some implications of the results are: As noted previously for the Minkowski case~\cite{bishop2024heating}, for small values of $r$ the heating effect is much stronger at the equator (where an accretion disk would be located) than at the poles. Further, the effect is enhanced when using a Schwarzschild background with a temperature increase on the equator at the ISCO of over $10^{12}$K predicted (which is much larger than the ambient temperature of an accretion disk of about $10^6$K). This would be sufficient to generate a gamma-ray burst as may have been observed for GW150914~\cite{connaughton2016fermi}.
\section{Conclusion}
\label{sec:con}
In this essay we discussed our recent results on the interactions of gravitational waves with viscous matter in different backgrounds. Generalizing away from a Minkowski background has meant that perturbed quantities no longer have simple forms in terms of elementary functions, and the metric perturbations, and the consequent GW damping and heating effects, are evaluated numerically.

Use of a general spherically symmetric, static background, rather than Minkowski or Schwarzschild, is more appropriate for many astrophysical problems, and we have applied the theory to core collapse supernovae, the post-merger stage of a binary neutron star merger, as well as to the effect of an accretion disk around merging black holes. In all cases, we found that the GW damping and heating effects were larger than in the case of a Minkowski background, and in some cases substantially larger. Generally, the effects are sufficiently large as to be astrophysically significant for all three scenarios considered. However, if the viscosity is at the bottom of the feasible range, and the frequency at the top of the range, then the astrophysical significance would be only marginal for CCSNe and BNS.

It is important to note that the accuracy of the predictions of GW damping and temperature increases are limited by two factors. Firstly, while the background used is more realistic than Minkowski, it is still a crude approximation since the spacetimes being modelled are highly dynamic, and further the GWs cannot properly be treated as small perturbations. Secondly, the temperature increases have taken account of only GW heating, and it can be expected that a more realistic model would provide a limit to temperature changes.

	
\newpage
\section*{Acknowledgements}
This work was supported by the National Research Foundation, South Africa, under Grants No. CPRR240314209194 and No. RA22111872966.
UT is supported in part by NSFC under Grant No. 11847239. VK is supported by the Kempe Foundation grant JCSMK24-507.

\pagestyle{empty}

\bibliographystyle{unsrt}

\begin{thebibliography}{10}

\bibitem{hawking1966perturbations}
Stephen~W Hawking.
\newblock Perturbations of an expanding universe.
\newblock {\em Astrophysical Journal, vol. 145, p. 544}, 145:544, 1966.

\bibitem{Bondi:1962px}
H.~Bondi, M.~G.~J. van~der Burg, and A.~W.~K. Metzner.
\newblock {Gravitational waves in general relativity. 7. Waves from
  axisymmetric isolated systems}.
\newblock {\em Proc. Roy. Soc. Lond. A}, 269:21--52, 1962.

\bibitem{Sachs:1962wk}
R.~K. Sachs.
\newblock {Gravitational waves in general relativity. 8. Waves in
  asymptotically flat space-times}.
\newblock {\em Proc. Roy. Soc. Lond. A}, 270:103--126, 1962.

\bibitem{bishop2022effect}
Nigel~T Bishop, Petrus~J van~der Walt, and Monos Naidoo.
\newblock Effect of a viscous fluid shell on the propagation of gravitational
  waves.
\newblock {\em Physical Review D}, 106(8):084018, 2022.

\bibitem{Naidoo:2021rzw}
Monos Naidoo, Nigel~T. Bishop, and Petrus~J. van~der Walt.
\newblock {Modifications to the signal from a gravitational wave event due to a
  surrounding shell of matter}.
\newblock {\em Gen. Rel. Grav.}, 53(8):77, 2021.

\bibitem{Bishop:2024htv}
Nigel~T. Bishop.
\newblock {Interaction between gravitational waves and a viscous fluid shell on
  a Schwarzschild background}.
\newblock {\em Phys. Rev. D}, 110(10):104062, 2024.

\bibitem{Bishop2026spherical}
Nigel~T. Bishop, Vishnu Kakkat, and Monos Naidoo.
\newblock {Gravitational wave interactions with a viscous fluid: Core collapse
  supernova, binary neutron star merger, and accretion around a black hole
  merger}.
\newblock {\em Phys. Rev. D}, in press, 2026.

\bibitem{connaughton2016fermi}
Valerie Connaughton, E~Burns, A~Goldstein, L~Blackburn, MS~Briggs, B-B Zhang,
  J~Camp, N~Christensen, CM~Hui, P~Jenke, et~al.
\newblock Fermi gbm observations of ligo gravitational-wave event gw150914.
\newblock {\em The Astrophysical Journal Letters}, 826(1):L6, 2016.

\bibitem{Bishop2024}
Nigel~T. Bishop, Vishnu Kakkat, Amos~S. Kubeka, Monos Naidoo, and Petrus~J.
  van~der Walt.
\newblock {The interaction of gravitational waves with matter}.
\newblock {\em Int. J. Mod. Phys. D}, 33(15):2441012, 2024.

\bibitem{Bishop:2004ug}
Nigel~T. Bishop.
\newblock {Linearized solutions of the Einstein equations within a Bondi-Sachs
  framework, and implications for boundary conditions in numerical
  simulations}.
\newblock {\em Class. Quant. Grav.}, 22:2393--2406, 2005.

\bibitem{abbott2016observation}
Benjamin~P Abbott, Richard Abbott, Thomas~D Abbott, Matthew~R Abernathy, Fausto
  Acernese, Kendall Ackley, Carl Adams, Thomas Adams, Paolo Addesso, Rana~X
  Adhikari, et~al.
\newblock Observation of gravitational waves from a binary black hole merger.
\newblock {\em Physical review letters}, 116(6):061102, 2016.

\bibitem{Bethe:1990mw}
H.~A. Bethe.
\newblock {Supernova mechanisms}.
\newblock {\em Rev. Mod. Phys.}, 62:801--866, 1990.

\bibitem{Janka:2012wk}
Hans-Thomas Janka.
\newblock {Explosion Mechanisms of Core-Collapse Supernovae}.
\newblock {\em Ann. Rev. Nucl. Part. Sci.}, 62:407--451, 2012.

\bibitem{Jerkstrand:2025bea}
Anders Jerkstrand, Dan Milisavljevic, and Bernhard M{\"u}ller.
\newblock Core-collapse supernovae.
\newblock {\em arXiv preprint arXiv:2503.01321}, 2025.

\bibitem{Yakunin:2017tus}
Konstantin~N. Yakunin, Anthony Mezzacappa, Pedro Marronetti, Eric~J. Lentz,
  Stephen~W. Bruenn, W.~Raphael Hix, O.~E. Bronson~Messer, Eirik Endeve,
  John~M. Blondin, and J.~Austin Harris.
\newblock {The Gravitational Wave Signal of a Core Collapse Supernova Explosion
  of a 15 M$_\odot$ Star}, 1 2017.

\bibitem{Andresen:2016pdt}
H.~Andresen, Bernhard M\"uller, Ewald M\"uller, and Hans-Thomas Janka.
\newblock {Gravitational Wave Signals from 3D Neutrino Hydrodynamics
  Simulations of Core-Collapse Supernovae}.
\newblock {\em Mon. Not. Roy. Astron. Soc.}, 468(2):2032--2051, 2017.

\bibitem{Muller:2020ard}
B.~M\"uller.
\newblock {Hydrodynamics of core-collapse supernovae and their progenitors}.
\newblock {\em Astrophysics}, 6:3, 2020.

\bibitem{Ott:2012jq}
C.~D. Ott, E.~P. O'Connor, S.~Gossan, E.~Abdikamalov, U.~C.~T. Gamma, and
  S.~Drasco.
\newblock {Core-Collapse Supernovae, Neutrinos, and Gravitational Waves}.
\newblock {\em Nucl. Phys. B Proc. Suppl.}, 235-236:381--387, 2013.

\bibitem{Kuroda:2016bjd}
Takami Kuroda, Kei Kotake, and Tomoya Takiwaki.
\newblock {A new Gravitational-wave Signature From Standing Accretion Shock
  Instability in Supernovae}.
\newblock {\em Astrophys. J. Lett.}, 829(1):L14, 2016.

\bibitem{Andresen:2018aom}
H.~Andresen, E.~M\"uller, H.Th. Janka, A.~Summa, K.~Gill, and M.~Zanolin.
\newblock {Gravitational waves from 3D core-collapse supernova models: The
  impact of moderate progenitor rotation}.
\newblock {\em Mon. Not. Roy. Astron. Soc.}, 486(2):2238--2253, 2019.

\bibitem{boccioli2024physics}
Luca Boccioli and Lorenzo Roberti.
\newblock The physics of core-collapse supernovae: Explosion mechanism and
  explosive nucleosynthesis.
\newblock {\em Universe}, 10(3):148, 2024.

\bibitem{abbott2017gw170817}
Benjamin~P Abbott, Rich Abbott, Thomas~D Abbott, Fausto Acernese, Kendall
  Ackley, Carl Adams, Thomas Adams, Paolo Addesso, Rana~X Adhikari, Vaishali~B
  Adya, et~al.
\newblock Gw170817: observation of gravitational waves from a binary neutron
  star inspiral.
\newblock {\em Physical review letters}, 119(16):161101, 2017.

\bibitem{abbott2017multi}
BP~Abbott, R~Abbott, TD~Abbott, F~Acernese, K~Ackley, C~Adams, T~Adams,
  P~Addesso, RX~Adhikari, VB~Adya, et~al.
\newblock Multi-messenger observations of a binary neutron star merger** any
  correspondence should be addressed to lvc. publications@ ligo. org.
\newblock {\em The Astrophysical Journal Letters}, 848(2):l12, 2017.

\bibitem{abbott2020gw190425}
Benjamin~P Abbott, Robert Abbott, TD~Abbott, S~Abraham, Fausto Acernese,
  K~Ackley, C~Adams, RX~Adhikari, VB~Adya, Christoph Affeldt, et~al.
\newblock Gw190425: Observation of a compact binary coalescence with total
  mass~ \(3.4 m\odot\).
\newblock {\em The Astrophysical Journal}, 892(1):L3, 2020.

\bibitem{abbott2017gravitational}
Benjamin~P Abbott, Robert Abbott, TD~Abbott, F~Acernese, K~Ackley, C~Adams,
  T~Adams, P~Addesso, RX~Adhikari, VB~Adya, et~al.
\newblock Gravitational waves and gamma-rays from a binary neutron star merger:
  Gw170817 and grb 170817a.
\newblock {\em The Astrophysical Journal Letters}, 848(2):L13, 2017.

\bibitem{hammond2025investigating}
P~Hammond, A~Clevinger, M~Albino, V~Dexheimer, S~Bernuzzi, C~Brown, W~Cook,
  B~Daszuta, J~Fields, E~Grundy, et~al.
\newblock Investigating the impact of higher-order phase transitions in binary
  neutron-star mergers.
\newblock {\em arXiv preprint arXiv:2508.10698}, 2025.

\bibitem{fonseca2021refined}
Emmanuel Fonseca, H~Thankful Cromartie, Timothy~T Pennucci, Paul~S Ray, A~Yu
  Kirichenko, Scott~M Ransom, Paul~B Demorest, Ingrid~H Stairs, Zaven
  Arzoumanian, Lucas Guillemot, et~al.
\newblock Refined mass and geometric measurements of the high-mass psr j0740+
  6620.
\newblock {\em The Astrophysical Journal Letters}, 915(1):L12, 2021.

\bibitem{antoniadis2013massive}
John Antoniadis, Paulo~CC Freire, Norbert Wex, Thomas~M Tauris, Ryan~S Lynch,
  Marten~H Van~Kerkwijk, Michael Kramer, Cees Bassa, Vik~S Dhillon, Thomas
  Driebe, et~al.
\newblock A massive pulsar in a compact relativistic binary.
\newblock {\em Science}, 340(6131):1233232, 2013.

\bibitem{fontbute2025gravitational}
Joan Fontbut{\'e}, Sebastiano Bernuzzi, Piero Rettegno, Simone Albanesi, and
  Wolfgang Tichy.
\newblock Gravitational scattering of two neutron stars.
\newblock {\em arXiv preprint arXiv:2506.11204}, 2025.

\bibitem{abbott2019properties}
BPea Abbott, R~Abbott, TD~Abbott, F~Acernese, K~Ackley, C~Adams, T~Adams,
  P~Addesso, RX~Adhikari, VB~Adya, et~al.
\newblock Properties of the binary neutron star merger gw170817.
\newblock {\em Physical Review X}, 9(1):011001, 2019.

\bibitem{raithel2021realistic}
Carolyn~A Raithel, Vasileios Paschalidis, and Feryal {\"O}zel.
\newblock Realistic finite-temperature effects in neutron star merger
  simulations.
\newblock {\em Physical Review D}, 104(6):063016, 2021.

\bibitem{baiotti2017binary}
Luca Baiotti and Luciano Rezzolla.
\newblock Binary neutron star mergers: a review of einstein’s richest
  laboratory.
\newblock {\em Reports on Progress in Physics}, 80(9):096901, 2017.

\bibitem{paschalidis2017rotating}
Vasileios Paschalidis and Nikolaos Stergioulas.
\newblock Rotating stars in relativity.
\newblock {\em Living Reviews in Relativity}, 20(1):7, 2017.

\bibitem{bishop2024heating}
Vishnu Kakkat, Nigel~T Bishop, and Amos~S. Kubeka.
\newblock Gravitational wave heating.
\newblock {\em Physical Review D}, 109:024013, 2024.

\bibitem{scientific2016tests}
LIGO Scientific, Virgo Collaborations, BP~Abbott, R~Abbott, TD~Abbott,
  MR~Abernathy, F~Acernese, K~Ackley, C~Adams, T~Adams, et~al.
\newblock Tests of general relativity with gw150914.
\newblock {\em Physical review letters}, 116(22):221101, 2016.

\bibitem{shakura1976theory}
NI~Shakura and RA~Sunyaev.
\newblock A theory of the instability of disk accretion on to black holes and
  the variability of binary x-ray sources, galactic nuclei and quasars.
\newblock {\em Monthly Notices of the Royal Astronomical Society},
  175(3):613--632, 1976.

\bibitem{arai1995accretion}
K~Arai and M~Hashimoto.
\newblock Accretion disk models with hydrogen burning around a black hole.
\newblock {\em Astronomy and Astrophysics, v. 302, p. 99}, 302:99, 1995.

\bibitem{abramowicz2013foundations}
Marek~A Abramowicz and P~Chris Fragile.
\newblock Foundations of black hole accretion disk theory.
\newblock {\em Living Reviews in Relativity}, 16:1--88, 2013.

\bibitem{rembiasz2017measurements}
Tomasz Rembiasz, Martin Obergaulinger, Pablo Cerd{\'a}-Dur{\'a}n,
  Miguel-{\'A}ngel Aloy, and Ewald M{\"u}ller.
\newblock On the measurements of numerical viscosity and resistivity in
  eulerian mhd codes.
\newblock {\em The Astrophysical Journal Supplement Series}, 230(2):18, 2017.

\bibitem{zappa2023binary}
Francesco Zappa, Sebastiano Bernuzzi, David Radice, and Albino Perego.
\newblock Binary neutron star merger simulations with neutrino transport and
  turbulent viscosity: impact of different schemes and grid resolution.
\newblock {\em Monthly Notices of the Royal Astronomical Society},
  520(1):1481--1503, 2023.

\end{thebibliography}

\end{document}